\documentclass[hyper,11pt,letterpaper]{JHEP3}

\usepackage{epsfig}       

\usepackage{amsfonts}     
\usepackage{alltt}        

\title{\texttt{Vscape V1.1.0}\\
An Interactive Tool for Metastable Vacua}  
\author{Korneel van den Broek \\
        \\
        Department of Physics and NHETC \\
        Rutgers University \\
        Piscataway, NJ 08854 \\
        E-mail: \email{korneel@physics.rutgers.edu}}
\abstract{
\texttt{Vscape} is an interactive tool for studying the one-loop effective potential of an ungauged supersymmetric model of chiral multiplets. The program allows the user to define a supersymmetric model by specifying the superpotential. The F-terms and the scalar and fermionic mass matrices are calculated symbolically. The program then allows you to search numerically for (meta)stable minima of the one-loop effective potential. Additional commands enable you to further study specific minima, by e.g. computing the mass spectrum for those vacua. \texttt{Vscape} combines the flexibility of symbolic software, with the speed of a numerical package.}

\preprint{RUNHETC-06-07}

\begin{document}

\section{Introduction}
The requirement that supersymmetry is broken at the global minima in field theories with dynamical supersymmetry breaking\cite{ISW} puts tight constraints on model building. Recently, dynamical supersymmetry breaking in metastable vacua has attracted attention as a promising phenomenological possibility. This greatly simplifies the model building problems. The metastable states are viable candidates to describe our world, when the tunneling probability to the supersymmetric ground state is highly suppressed. As argued in \cite{ISS1}, metastable vacua might be quite generic in supersymmetric theories\cite{Retrofit}\cite{DJohn}. With some mild assumptions, low energy supersymmetry breaking even requires our world as we know it to be a long lived metastable state\cite{ISS2}. From the string theory point of view, non-supersymmetric flux compactifications might also be rather generic in the landscape of vacua\cite{DenefDouglasEtAl}. 

An important role in studying the moduli space of supersymmetric theories is played by the one-loop effective potential of the field theory. The classical (pseudo)moduli space can contain flat directions. The one-loop corrections often lift those pseudomoduli directions of the classical theory, thus creating isolated (meta)stable vacuum. Generically, several fields will have vacuum expectation values (vevs) at those local minima. The important phenomenological objective is then to create models that communicate this dynamical breaking of supersymmetry to the supersymmetric standard model such that one gets attractive superpartner spectra\cite{GuidiceRattazziEtal}. 

The computation of the Coleman-Weinberg potential\cite{CW} in realistic models often becomes rather involved as the mass matrices become large and one can expect several fields to gain vevs. Even for symbolic software package, multiple diagonalization of large mass matrices is rather time consuming, while numerical packages usually do not provide the flexibility of working with symbolic formulas. \verb|Vscape| aims to bridge this gap between symbolic and numerical packages. The flexibility of the \verb|Vscape| users interface gives the user the freedom to define the physics of the model symbolically, while the computationally intensive evaluations of the Coleman-Weinberg potential rely on fast numerical C++ routines. 

In addition \verb|Vscape| provides the necessary tools to analyse a metastable vacuum once one is found. Commands are provided to compute the mass spectrum at the local minimum. Those results can then be used as input data for a spectrum generator such as \verb|SOFTSUSY|\cite{SOFTSUSY}. Vscape has a command implemented with allows for the creation of files formatted according the Model Input File standard described in the Les Houches Accord 1.0. to supply the high energy input to the spectrum generator.

The code for \verb|Vscape| was written in object-oriented C++ since this language compiles to speed efficient programs, allowing for fast diagonalization of large matrices for the Coleman-Weinberg potential. Using C++ also has the advantage that an extensive collection very powerful, mathematical libraries are freely available. \verb|Vscape| makes use of the recent libraries GSL \cite{GSL}, GiNaC and CLN \cite{GiNaC}. \verb|Vscape| also has Tab-completion build in, if one has the readline library installed. The program can freely be downloaded and installed from either:
\begin{alltt}
   \href{http://www.physics.rutgers.edu/~korneel/vscape/vscape.html}{http://www.physics.rutgers.edu/~korneel/vscape/vscape.html}
\end{alltt}
or 
\begin{alltt}
   \href{http://projects.hepforge.org/vscape}{http://projects.hepforge.org/vscape/}
\end{alltt}
Detailed installation instruction (Linux and Windows) can also be found there. Updates and fixes of possible errors will also appear on those websites.

In section \ref{SectionPhysics}, we briefly introduce the physics behind the commands of \verb|Vscape|. We then give a short account on how the program is structured in section \ref{SectionStructure}. Section \ref{SectionOutlook} lists several possibilities for further extensions of the program. Appendix \ref{GeneralCommands} and \ref{SpecificCommands} contain the detailed syntax and functionality of all the commands understood by \verb|Vscape|. Section \ref{SubSectionPrecision} discusses the precision of the numerical computations, while section \ref{SubSectionControlParameters} lists the commands to change the control parameters of the program to influence the algorithms underlying the program.

\section{Physics Overview} \label{SectionPhysics}

\subsection{The supersymmetric model of chiral fields}
The current version of \verb|Vscape| allows one to study various $\mathcal{N} = 1$ supersymmetric models with chiral fields\cite{Martin}. The user defines the model by specifying the parameters and chiral fields $\phi_i$ of the model. In addition, one has to specify the subset of fields which are a priori allowed to obtain a vev for their scalar component. Let us denote this subset of background fields as $\varphi_i$, while we will use the symbol $\tilde{\varphi}_n$ for the fields that are not allowed to get a scalar vev. The model is then defined through the superpotential $W$ and the D-term potential $V_D$. The D-term potential is only taken into account in the tree level potential\footnote{We include the D-terms in the tree level potential to accomodate models like \cite{LeanMeanPentagon} where the hidden sector itself does not have a D-term potential but where the Supersymmetric Standard Model (SSM) sector does contribute D-terms to the tree level potential. This also allows us to described the metastable vacua of the models studied in \cite{ISS1} where the theory is gauged, but the effect of the gauge fields drops out of the one-loop correction.}. The current version of the program does not include terms from vector multiplets in the Coleman-Weinberg potential computation. The program assumes a canonical K\"ahler potential,
\begin{equation}
  K = \mathrm{Tr} \, \phi^{\dagger} \phi \, .
\end{equation}

When a new model is loaded, the program computes and stores the following information symbolically:
\begin{eqnarray}
  V_0      & = &   \left. W^*_i \right|_{\tilde{\varphi}_n = 0} 
                   \left. W^i   \right|_{\tilde{\varphi}_n = 0} 
                 + \left. V_D   \right|_{\tilde{\varphi}_n = 0}                      \\
  Fterms_i & = &   \left. W_i   \right|_{\tilde{\varphi}_n = 0}                      \\
  mF_{ij}  & = &   \left. W_{ij} \right|_{\tilde{\varphi}_n = 0}                     \\
  mB_{ij}  & = &   \left. W^*_{ijk} W^k \right|_{\tilde{\varphi}_n = 0}
\end{eqnarray}
where an index $i$ to the superpotential $W$ stands for a derivative with respect to field $\phi_i$. The $|_{\tilde{\varphi}_n = 0}$ indicates that the fields that are not allowed to get a vev by the user, are set to zero in the expression. The commands detailed in section \ref{SymbolicCommands} give you access to this information.

\subsection{Coleman-Weinberg computation}
To compute the Coleman-Weinberg potential, the numerical values for the vevs and parameters are substitued into the matrices $mF$ and $mB$. The program then constructs the numerical mass squared matrices
\begin{eqnarray} 
  M^{\dagger}_{1/2}M_{1/2}
            &=& W^*_{jk} W^{ik}                              \label{FermionMassMatrixSquared}       \\
  M^2_0     &=& \left( \begin{array}{cc}
                      W^*_{jk} W^{ik}   &   W^*_{ijk} W^k    \\
                      W^{ijk} W^*_k     &   W^*_{ik} W^{jk} 
                \end{array} \right)                          \label{BosonMassMatrixSquared} 
\end{eqnarray}
The mass squared eigenvalues of these matrices, $(m^2_{1/2})_i$ and $(m^2_0)_i$ respectively, are computed and substituted in the Coleman-Weinberg formula\cite{CW}
\begin{equation}
  V_{\mathrm{cw}} = \frac{1}{64 \pi^2} \left(
                           \sum_i (m^2_0)^2_i     \ln (m^2_0)_i 
                       - 2 \sum_i (m^2_{1/2})^2_i \ln (m^2_{1/2})_i 
                    \right)  \, .
\end{equation}
The effective potential is thus a function of the vevs of the background fields $\varphi_i$. 

To check the proper functioning of the Coleman-Weinberg computation, \verb|Vscape| allows to check the supertrace of the tree-level squared mass matrices. The supertrace is computed completely analogous to the Coleman-Weinberg potential, in that the eigenvalues of the matrices (\ref{FermionMassMatrixSquared}) and (\ref{BosonMassMatrixSquared}) are computed. The eigenvalues are then substituted in the formula:
\begin{equation}
  \mathrm{STr}(M^2) =     \sum_i (m^2_0)_i 
                      - 2 \sum_i (m^2_{1/2})_i   \, .
\end{equation}

\subsection{Metastable vacua}
\verb|Vscape| contains several commands to scan for local minima on the one-loop effective potential surface. Once a local minima of the effective potential is found, one can compute the one-loop corrected masses at the minima from:
\begin{equation}
  \left. V_{\mathrm{eff}} \right|_{\mathrm{min}} 
    =   \left. V_{\mathrm{eff}} \right|_{\mathrm{min}} 
      + \frac{1}{2} \left. H_{ij} \right|_{\mathrm{min}} \varphi_i \varphi_j 
      + \ldots
\end{equation}
where $H_{ij}$ is the Hessian matrix. The eigenvalues of the Hessian are the mass squareds of the associated mass eigenstates. The mass squared eigenvalues allow you to verify (within error bounds, see section \ref{SubSectionPrecision}) whether the minimum is indeed stable. Depending on the setup studied, the actual mass squared eigenvalues can be used as input parameters for SSM spectrum generators.

\section{Structure of the program} \label{SectionStructure}
The physics of the supersymmetric model is entirely encoded in the class CModel which is defined in the files \verb|phenomodel.h| and \verb|phenomodel.cc|. CModel has several routines relying on symbolic computations using the GiNaC library while other GSL-based routines are numerical.

The user-interface of \verb|Vscape| was build on \verb|ginsh|, the interactive frontend for the GiNaC symbolic computation framework \cite{GiNaC}. This lightweight package for symbolic computations provided the appropriate user-interface to communicate between the phenomenological model and the user. We slightly extended the commands of the original \verb|ginsh| application, to fit our needs. 

\verb|Ginsh| is implemented using the tools flex and bison which provide the detailed code for the lexer which reads the input (\verb|ginsh_lexer.ll|) and parser which interprets the input (\verb|ginsh_parser.yy|). \verb|Ginsh| contains several commands, which are not related to the underlying phenomenological model. Those commands are detailed in Appendix \ref{GeneralCommands}.

The original \verb|ginsh| implementation allowed for additional commands to be added by other programmers in the separate file \verb|ginsh_extensions.h|. The commands that specifically relate to the phenomenological model are defined in \verb|vscape.h| and \verb|vscape_functions.h|. They are included via \verb|ginsh_extensions.h|. The shorter functions are implemented in \verb|vscape.h| while the longer, mathematically involved algorithms such as minimization, are implemented in \verb|vscape_functions.h|. Appendix \ref{SpecificCommands} details the specific commands which allow you to interact with the phenomenological model CModel.

\section{Outlook} \label{SectionOutlook}
This project grew out of a program written to study the Pentagon Model \cite{LeanMeanPentagon} \cite{BvdB}. Originally, the definition of the model was hardwired into the code. The added symbolic interface of \verb|Vscape| gives the user the freedom to define the physics of the model her/himself. The commands available in \verb|Vscape| provide the functionality that we needed to study the Pentagon Model. There are several additional features which might be interesting to include.

\begin{itemize}
\item Including vector multiplets in the computation of the Coleman-Weinberg potential. 

\item Allowing for a more general K\"ahler potential instead of the currently assumed canonical one. 

\item Specific commands to study the barrier heights and associated tunneling probabilities.

\item Extending the user interface by introducing indexed symbols and control flow statements such as if-then statements and loops.

\item Faster symbolic-numeric interface (technical extension). Currently the interface between the symbolic library GiNaC and the numeric library GSL is based on the GiNaC subs() command, which is rather time consuming. One could create a derived class of the GiNaC numeric class, which stores numerical data as actual C++ doubles. The implementation of the automatically invoked eval() command in the class should then take care of numeric evaluation, without having to do explicit subs() command. GSL could then manipulate directly on the doubles of this class.

\end{itemize}

\section{Acknowledgements}
I would like to thank my advisor T.~Banks for his support and guidance throughout all the stages of this project. I am very grateful to R.~Essig, K.~Sinha, S.~Thomas and G.~Torroba for testing the program and for their suggestions. I also benefited from discussions with J.~Jones, S.~Lukic and J.~Mason. A word of thanks to all the programmers contributing to impressive libraries that are freely available for everyone. I would like to thank SCIPP, where part of this work was done, for their hospitality. This work is supported by the Rutgers Department of Physics and Astronomy and DOE grant number DE-FG02-96ER40949.

\appendix
\section{General Symbolic Commands} \label{GeneralCommands}
In this section, we detail the general math commands known to the Vscape. It is a slightly extended set of the original ginsh commands. The original manual of ginsh can be found as the linux ginsh manual.

\subsection{Running Vscape}
You run Vscape by typing
\begin{alltt}
   ./Vscape [file ...]
\end{alltt}  
where the list of files is optional. Vscape will first attempt to read and interpret the commands from the file(s). When no files were specified or all the commands in the files have been executed, Vscape displays a prompt (\verb|> |) signifying that it is ready to accept your input. Acceptable input are numeric or symbolic expressions consisting of numbers (e.g. \verb|42|, \verb|2/3| or \verb|0.17|), symbols (e.g. \verb|x| or \verb|result|), mathematical operators like \verb|+| and \verb|*|, and functions (e.g. \verb|sin| or \verb|normal|). Every input expression must be terminated with either a semicolon (\verb|;|) or a colon (\verb|:|). If terminated with a semicolon, Vscape will evaluate the expression and print the result to stdout. If terminated with a colon, Vscape will only evaluate the expression but not print the result. It is possible to enter multiple expressions on one line. Whitespace (spaces, tabs, newlines) can be applied freely between tokens. To quit Vscape, enter \verb|quit| or \verb|exit|, or type an EOF (Ctrl-D) at the prompt.

\subsection{Comments}
Anything following a double slash (\verb|//|) up
to the end of the line, and all lines starting with a hash mark
(\verb|#|) are treated as a comment and ignored.

\subsection{Numbers}
Vscape accepts numbers in the usual decimal
notations. This includes arbitrary precision integers and rationals
as well as floating point numbers in standard or scientific
notation (e.g. \verb|1.2E6|). The general rule is that if a number
contains a decimal point (\verb|.|), it is an (inexact) floating
point number, otherwise it is an (exact) integer or rational.
Integers can be specified in binary, octal, hexadecimal or
arbitrary (2-36) base by prefixing them with \verb|#b|, \verb|#o|,
\verb|#x|, or \verb|#|n\verb|R|, respectively.

\subsection{Symbols}
Symbols are made up of a string of alphanumeric
characters and the underscore (\verb|_|), with the first character
being non-numeric. E.g. \verb|a| and \verb|mu_1| are acceptable
symbol names, while \verb|2pi| is not. It is possible to use
symbols with the same names as functions (e.g. \verb|sin|), Vscape
is able to distinguish between the two.

Symbols can be assigned values by entering
\begin{alltt}
   \textrm{\textit{symbol}} = \textrm{\textit{expression}}
\end{alltt}
To unassign the value of an assigned symbol, type
\begin{alltt}
   unassign('\textrm{\textit{symbol}}');
\end{alltt}
Assigned symbols are automatically replaced by their assigned value when they are used. To refer to the unevaluated symbol, put single quotes (\verb|'|) around the name, as demonstrated for the \verb|unassign| command above.

Symbols are considered to be in the complex or real domain depending on the mode in which they were defined. If the program is in the complex mode, every new symbol that has not been used before during the session is considered to be complex. If the program is in the real mode, every newly encountered symbol is considered to be real. The mode of the program only affects newly defined symbols. Thus when the program is in the real mode, symbols that were defined previously in the complex mode, remain complex. This mode of the program is switch using the keywords \verb|real_symbols| and \verb|complex_symbols|. The program is in the complex mode by default.

The following symbols are pre-defined constants that cannot be assigned a value by the user:
\begin{alltt}
   Pi      \textrm{Archimedes' Constant}
   Catalan \textrm{Catalan's Constant}               
   Euler   \textrm{Euler-Mascheroni Constant} 
   I       \textrm{imaginary unit \(i\)}          
   FAIL    \textrm{an object of the GiNaC `fail' class}
\end{alltt}
There is also the special
\begin{verbatim}
   Digits
\end{verbatim}
symbol that controls the numeric precision of calculations with inexact numbers. Assigning an integer value to digits will change the precision to the given number of decimal places.

\subsection{Wildcards}
The \verb|has()|, \verb|find()|, \verb|match()| and \verb|subs()| functions accept wildcards as placeholders for expressions. These have the syntax
\begin{alltt}
   \$\textrm{\textit{number}}
\end{alltt}
for example \verb|$0|, \verb|$1| etc.

\subsection{Last printed expressions}
Vscape provides the three special symbols
\begin{alltt}
   \%, \%\% \textrm{\textit{and}} \%\%\%
\end{alltt}
that refer to the last, second last, and third last printed expression, respectively. These are handy if you want to use the results of previous computations in a new expression.

\subsection{Operators}
Vscape provides the following operators, listed in falling order of precedence:
\begin{alltt}
   !  \textrm{postfix factorial}
   ^  \textrm{powering} 
   +  \textrm{unary plus}
   -  \textrm{unary minus} 
   *  \textrm{multiplication}
   /  \textrm{division}
   +  \textrm{addition}
   -  \textrm{subtraction}
   <  \textrm{less than}
   >  \textrm{greater than}
   <= \textrm{less or equal} 
   >= \textrm{greater or equal}
   == \textrm{equal}
   != \textrm{not equal}
   =  \textrm{symbol assignment}
\end{alltt}
All binary operators are left-associative, with the exception of \verb|^| and \verb|=| which are right-associative. The result of the assignment operator (\verb|=|) is its right-hand side, so it's possible to assign multiple symbols in one expression (e.g. \verb|a = b = c = 2;|).

\subsection{Lists}
A list consists of an opening curly brace (\verb|{|), a (possibly empty) comma-separated sequence of expressions, and a closing curly brace (\verb|}|). A list is not a set in that it can contain several times the same element (e.g. \verb|{1,2,3,4,3,3,x^2}|).

\subsection{Matrices}
A matrix consists of an opening square bracket (\verb|[|), a non-empty comma-separated sequence of matrix rows, and a closing square bracket (\verb|]|). Each matrix row consists of an opening square bracket (\verb|[|), a non-empty
comma-separated sequence of expressions, and a closing square bracket (\verb|]|). If the rows of a matrix are not of the same length, the width of the matrix becomes that of the longest row and shorter rows are filled up at the end with elements of value zero.

\subsection{Functions}
A function call in Vscape has the form 
\begin{alltt}
   \textrm{\textit{name}}(\textrm{\textit{arguments}})
\end{alltt}
where \textit{arguments} is a comma-separated sequence of expressions. Vscape provides a couple of built-in functions and also `imports' all symbolic functions defined by GiNaC and additional libraries. There is no way to define your own functions other than linking Vscape against a library that defines symbolic GiNaC functions. 

Vscape provides Tab-completion on function names: if you type the first part of a function name, hitting Tab will complete the name if possible. If the part you typed is not unique, hitting Tab again will display a list of matching functions. Hitting Tab twice at the prompt will display the list of all available functions. 

A list of the built-in functions follows. They nearly all work as the respective GiNaC methods of the same name, so I will not describe them in detail here. Please refer to the GiNaC documentation.
\newlength{\mylength}                 
\newlength{\myindent}
\setlength{\mylength}{\linewidth}     
\settowidth{\myindent}{\tt XXXXXX}    
\addtolength{\mylength}{-1\myindent}  
\begin{alltt}
   append(\textrm{\textit{list1}}, \textrm{\textit{list2}})
      \textrm{appends list2 to list1}
   charpoly(\textrm{\textit{matrix}}, \textrm{\textit{symbol}}) 
      \textrm{characteristic polynomial of a matrix}
   coeff(\textrm{\textit{expression}}, \textrm{\textit{object}}, \textrm{\textit{number}}) 
      \textrm{extracts coefficient of \textit{object}}^\textrm{\textit{number} from a polynomial}
   collect(\textrm{\textit{expression}}, \textrm{\textit{object-or-list}})
      \textrm{collects coefficients of like powers (result in recursive form)}
   collect_distributed(\textrm{\textit{expression}}, \textrm{\textit{list}})
      \textrm{collects coefficients of like powers (result in distributed form)}
   collect_common_factors(\textrm{\textit{expression}})
      \textrm{collects common factors from the terms of sums}
   conjugate(\textrm{\textit{expression}})
      \textrm{complex conjugation}
   content(\textrm{\textit{expression}}, \textrm{\textit{symbol}})
      \textrm{content part of a polynomial}
   decomp_rational(\textrm{\textit{expression}}, \textrm{\textit{symbol}})
      \textrm{decompose rational function into polynomial and proper rational function}
   degree(\textrm{\textit{expression}}, \textrm{\textit{object}})
      \textrm{degree of a polynomial}
   denom(\textrm{\textit{expression}})
      \textrm{denominator of a rational function}
   determinant(\textrm{\textit{matrix}})
      \textrm{determinant of a matrix}
   diag(\textrm{\textit{expression...}})
      \textrm{constructs diagonal matrix}
   diff(\textrm{\textit{expression}}, \textrm{\textit{symbol}} [, \textrm{\textit{number}}])
      \textrm{partial differentiation}
   divide(\textrm{\textit{expression}}, \textrm{\textit{expression}})
      \textrm{exact polynomial division}
   eigenherm(\textrm{\textit{matrix}})
      \parbox[t]{\mylength}{\textrm{compute the eigenvalues and eigenvectors of an n by n hermitian matrix M of numbers. The function returns a list with a 1 by n matrix of eigenvalues followed by the n by n unitary matrix were each column is an eigenvector. The columns of eigenvector are ordered in the same way the eigenvalues appear in the 1 by n matrix. Note: this is a numerical command, it does not accept expressions that cannot be evaluated numerically. The command works with double precision internally.}}
\vspace{.02in}   
   eigensymm(\textrm{\textit{matrix}})
      \parbox[t]{\mylength}{\textrm{compute the eigenvalues and eigenvectors of an n by n real symmetric matrix M of numbers. The function returns a list with a 1 by n matrix of eigenvalues followed by the n by n orthogonal matrix were each column is an eigenvector. The columns of eigenvector are ordered in the same way the eigenvalues appear in the 1 by n matrix. Note: this is a numerical command, it does not accept expressions that cannot be evaluated numerically. The command works with double precision internally.}}
\vspace{.02in}   eval(\textrm{\textit{expression}} [, \textrm{\textit{level}}])
      \textrm{evaluates an expression, replacing symbols by their assigned value}
   evalf(\textrm{\textit{expression}} [, \textrm{\textit{level}}])
      \textrm{evaluates an expression to a floating point number}
   evalm(\textrm{\textit{expression}})
      \textrm{evaluates sums, products and integer powers of matrices}
   expand(\textrm{\textit{expression}})
      \textrm{expands an expression}
   find(\textrm{\textit{expression}}, \textrm{\textit{pattern}})
      \textrm{returns a list of all occurrences of a pattern in an expression}
   fsolve(\textrm{\textit{expression}}, \textrm{\textit{symbol}}, \textrm{\textit{number}}, \textrm{\textit{number}})
      \textrm{numerically find rootof a real-valued function within an interval}
   gcd(\textrm{\textit{expression}}, \textrm{\textit{expression}})
      \textrm{greatest common divisor}
   has(\textrm{\textit{expression}}, \textrm{\textit{pattern}})
      \textrm{returns `1' if the first expression contains the pattern as a subexpression, `0' otherwise}
   importfile(\textrm{\textit{filename}})
      \textrm{read commands from \textit{filename}. Note that this command cannot be nested in files.}
   integer\_content(\textrm{\textit{expression}})
      \textrm{integer content f a polynomial}
   inverse(\textrm{\textit{matrix}})
      \textrm{inverse of a matrix}
   is(\textrm{\textit{relation}})
      \textrm{returns `1' if the relation is true, `0' otherwise (false or undecided)}
   lcm(\textrm{\textit{expression}}, \textrm{\textit{expression}})
      \textrm{least common multiple}
   lcoeff(\textrm{\textit{expression}}, \textrm{\textit{object}})
      \textrm{leading coefficient of a polynomial}
   ldegree(\textrm{\textit{expression}}, \textrm{\textit{object}})
      \textrm{low degree of a polynomial}
   lsolve(\textrm{\textit{equation-list}}, \textrm{\textit{symbol-list}})
      \textrm{solve system of linear equations}
   map(\textrm{\textit{expression}}, \textrm{\textit{pattern}})
      \parbox[t]{\mylength}{\textrm{apply function to each operand, the function to be applied is specified as a pattern with the `\$0' wildcard standing for all the operands}}
\vspace{.02in}   match(\textrm{\textit{expression}}, \textrm{\textit{pattern}})
      \textrm{check whether expression matches a pattern}
      \textrm{returns a list of wildcard substitutions or `FAIL' if there is no match}
   minus(\textrm{\textit{list1}}, \textrm{\textit{list2}})
      \textrm{returns the set-theoretic difference of list1 and list2}
   nops(\textrm{\textit{expression}})
      \textrm{number of operands in expression}
   normal(\textrm{\textit{expression}} [, \textrm{\textit{level}}])
      \textrm{rational function normalization}
   numer(\textrm{\textit{expression}})
      \textrm{numerator of a rational function}
   numer_denom(\textrm{\textit{expression}})
      \textrm{numerator and denumerator of a rational function as a list}
   op(\textrm{\textit{expression}}, \textrm{\textit{number}})
      \textrm{extract operand \textit{number} from \textit{expression}, can also be used to get elements in a matrix}
   power(\textrm{\textit{expression1}}, \textrm{\textit{expression2}})
      \textrm{exponentiation (equivalent to writing \textit{expression1}}^\textrm{\textit{expression2})}
   prem(\textrm{\textit{expression}}, \textrm{\textit{expression}}, \textrm{\textit{symbol}})
      \textrm{pseudo-remainder of polynomials}
   primpart(\textrm{\textit{expression}}, \textrm{\textit{symbol}})
      \textrm{primitive part of a polynomial}
   quo(\textrm{\textit{expression}}, \textrm{\textit{expression}}, \textrm{\textit{symbol}})
      \textrm{quotient of polynomials}
   rank(\textrm{\textit{matrix}})
      \textrm{rank of a matrix}
   rem(\textrm{\textit{expression}}, \textrm{\textit{expression}}, \textrm{\textit{symbol}})
      \textrm{remainder of polynomials}
   resultant(\textrm{\textit{expression}}, \textrm{\textit{expression}}, \textrm{\textit{symbol}})
      \textrm{resultant of two polynomials with respect to symbols}
   rnd(\textrm{\textit{number1}}, \textrm{\textit{number2}})
      \parbox[t]{\mylength}{\textrm{generate a random number with double precision with uniform distribution within the interval [number1, number2]}}
\vspace{.02in}   save(\textrm{\textit{filename}}[, \textrm{\textit{string}}] [, \textrm{\textit{expression}}])
      \textrm{open \textit{filename} and append \textit{string} and/or \textit{expression}; to file}
   series(\textrm{\textit{expression}}, \textrm{\textit{relation-or-symbol}}, \textrm{\textit{order}})
      \textrm{series expansion}
   sprem(\textrm{\textit{expression}}, \textrm{\textit{expression}}, \textrm{\textit{symbol}})
      \textrm{sparse pseudo-remainder of polynomials}
   sqrfree(\textrm{\textit{expression}} [, \textrm{\textit{symbol-list}}])
      \textrm{square-free factorization of a polynomial}
   sqrt(\textrm{\textit{expression}})
      \textrm{square root}
   symbolicmatrix(\textrm{\textit{integer1}}, \textrm{\textit{integer1}}, \textrm{\textit{symbol}})
   	  \textrm{create an \textit{integer1} by \textit{integer2} matrix with entries \textit{symbol\_{XX}}} 
   subs(\textrm{\textit{expression}}, \textrm{\textit{relation-or-list}}) 
   subs(\textrm{\textit{expression}}, \textrm{\textit{look-for-list}}, \textrm{\textit{replace-by-list}})
      \textrm{substitute subexpressions (you may use wildcards)}
   tcoeff(\textrm{\textit{expression}}, \textrm{\textit{object}})
      \textrm{trailing coefficient of a polynomial}
   time(\textrm{\textit{expression}})
      \textrm{returns the time in seconds needed to evaluate the given expression}
   trace(\textrm{\textit{matrix}})
      \textrm{trace of a matrix}
   transpose(\textrm{\textit{matrix}})
      \textrm{transpose of a matrix}
   unassign(\textrm{\textit{symbol}})
      \textrm{unassign an assigned symbol}
   unit(\textrm{\textit{expression}}, \textrm{\textit{symbol}})
      \textrm{unit part of a polynomial}
   unique(\textrm{\textit{list}})
      \textrm{removes all multiple occurrences of expressions in the list}
\end{alltt}

\subsection{Special Commands}
To exit Vscape, enter \verb|quit| or \verb|exit|. Vscape can display a (short) help for a given topic (mostly about functions and operators) by entering \verb|?|\textit{topic}. Typing \verb|??| will display a list of available help topics. The command
\begin{alltt}
   print(\textrm{\textit{expression}});
\end{alltt}
will print a dump
of GiNaC's internal representation for the given \textit{expression}.
This is useful for debugging and for learning about GiNaC
internals.
The command
\begin{alltt}
   print_latex(\textrm{\textit{expression}});
\end{alltt}
prints a LaTeX representation of the given \textit{expression}. The command
\begin{alltt}
   print_csrc(\textrm{\textit{expression}});
\end{alltt}
prints the given \textit{expression} in a way that can be used in a C or C++ program. The command
\begin{alltt}
   iprint(\textrm{\textit{expression}});
\end{alltt}
prints the given \textit{expression} (which must evaluate to an integer) in decimal, octal, and hexadecimal representations. Finally, the shell escape
\begin{alltt}
   ![\textrm{\textit{command}} [\textrm{\textit{arguments}}]]
\end{alltt}
passes the given \textit{command} and optionally \textit{arguments} to the shell for execution. With this method, you can execute shell commands from within Vscape without having to quit.

\subsection{Error messages}
When you enter something which Vscape is unable to parse, it will report
\begin{alltt}
   syntax error, unexpected \textrm{\textit{foo}} at \textrm{\textit{bar}}
\end{alltt}
Please check the syntax of your input and try again. If the computer reports
\begin{alltt}
   argument \textrm{\textit{num}} to \textrm{\textit{function}} must be a \textrm{\textit{type}}
\end{alltt}   
it means that the argument number \textit{num} to the given \textit{function} has to be of a certain type (e.g. a symbol, or a list). The first argument has number 0, the second argument number 1, etc.

\subsection{Examples}
\begin{verbatim}
   > a = x^2-x-2;
   -2-x+x^2
   > b = (x+1)^2;
   (x+1)^2
   > s = a/b;
   (1+x)^(-2)*(-2-x+x^2)
   > diff(s, x);
   -2*(1+x)^(-3)*(-2-x+x^2)+(-1+2*x)*(1+x)^(-2)
   > normal(s);
   (1+x)^(-1)*(-2+x)
   > x = 3^50;
   717897987691852588770249
   > s;
   717897987691852588770247/717897987691852588770250
   > Digits = 40;
   40
   > evalf(s);
   0.999999999999999999999995821133292704384960990679
   > unassign('x');
   x
   > s;
   (1+x)^(-2)*(-2-x+x^2)
   >
   > s = x^3 + x^2 + y + 1;
   1+y+x^3+x^2
   > map(s, $0^2);
   1+y^2+x^4+x^6
   > match(s, $0+$1^2);
   {$1==x,$0==1+y+x^3}
   >
   > series(sin(x),x==0,6);
   1*x+(-1/6)*x^3+1/120*x^5+Order(x^6)
   >
   > lsolve({3*x+5*y == 7}, {x, y});
   {x==-5/3*y+7/3,y==y}
   > lsolve({3*x+5*y == 7, -2*x+10*y == -5}, {x, y});
   {x==19/8,y==-1/40}
   >
   > M = [ [a, b], [c, d] ];
   [[-x+x^2-2,(x+1)^2],[c,d]]
   > op(M, 0); op(M, 2);
   -x+x^2-2
   c
   > determinant(M);
   -2*d-2*x*c-x^2*c-x*d+x^2*d-c
   > collect(%, x);
   (-d-2*c)*x+(d-c)*x^2-2*d-c
   >
   > save("test.txt", "# test saving M to file");
   Wrote to file <test.txt>: "# test saving M to file"
   > save("test.txt", "M = ", M);
   Wrote to file <test.txt>: "M = [[-2-x+x^2,(1+x)^2],[c,d]];"
   > save("test.txt", M);
   Wrote to file <test.txt>: "[[-2-x+x^2,(1+x)^2],[c,d]];"
   >
   > solve quantum field theory;
   syntax error, unexpected T_SYMBOL at quantum
   > quit
\end{verbatim}
In this example, three lines where added to the end of the file \verb|"test.txt"|: 
\begin{verbatim}
   # test saving M to file
   M =[[-2-x+x^2,(1+x)^2],[c,d]];
   [[-2-x+x^2,(1+x)^2],[c,d]];
\end{verbatim}
If the file did not exist originally, it was created the first time the command \verb|save()| was executed.

\section{Commands to find and analyse metastable vacua} \label{SpecificCommands}
In this section, we introduce the Vscape commands that allow you to interact with the build in supersymmetric model in Vscape.

\subsection{Defining the model}
First you need to define the supersymmetric model that you want to study. The command
\begin{alltt}
   ModelConstruct(\textrm{\textit{Params}}, \textrm{\textit{Fields}}, \textrm{\textit{Vevs}}, \textrm{\textit{W}}, \textrm{\textit{V\_D}})
\end{alltt}
allows you to specify the model. \textit{Params} is a list of parameters that appear in the model (e.g. coupling constants). \textit{Fields} is a list of expressions, every symbol in the list will be considered a field in your model. \textit{Vevs} is a subset of \textit{Fields}, it contains the fields which you allow to acquire a non-zero vacuum expectation value. The order in which the symbols appear in \textit{Fields} determines the ordering of fields in the mass matrices and in the list of F-terms (see example later on). The expression \textit{W}, specifies the superpotential and \textit{V\_D} is the expression for the D-term potential which is added to the tree level potential. Both expressions may only contain symbols that are listed either in \textit{Fields} or \textit{Params}. The command \verb|ModelConstruct()| will return \verb|0| if it successfully defined the model, otherwise it will show an error message. Once the model is successfully loaded, you can use all the other commands described below to study the model.

The command \verb|ModelConstruct()| computes the F-terms and mass-matrices symbolically. Depending on the number of fields in the model and calculational speed this command might take some time. Once the model is computed with \verb|ModelConstruct()|, you can use the command
\begin{alltt}
   ModelSave(\textrm{\textit{filename}})
\end{alltt}
to save the computed model to \textit{filename}. This command saves all the information about the model including F-terms and mass matrices such that, later on, when you want to load that specific model again, the computer does not have to redo the lengthy computation of the F-terms and mass matrices. \verb|ModelSave()| will return \verb|0| if it successfully saved the model otherwise it will show an error message. 

Reloading a specific model, is done by simply invoking \verb|importfile(|\textit{filename}\verb|)|. \verb|Vscape| will execute the commands in \textit{filename}. \textit{filename} was created by \verb|ModelSave()| such that it will first define the fields, parameters, vevs, superpotential, D-term potential, F-terms and mass matrices. The last line of \textit{filename} will call the command \verb|ModelLoad()| (discussed below) which does the actually loading of the model into \verb|Vscape|. So after invoking \verb|importfile(|\textit{filename}\verb|)|, where \textit{filename} is a file that was created with \verb|ModelSave()|, the model is loaded into \verb|Vscape| and you can use all the commands described below to study the model.

I would advise against manually editing files created by \verb|ModelSave()|, since they contain all the information of the model you defined with \verb|ModelConstruct()| in a consistent way. It contains the exact mass matrices and F-terms corresponding to the superpotential with the correct ordering. E.g. the order in which the F-terms appear in \textit{filename} is determined by the order in which the fields are listed in \textit{filename}. 

The command
\begin{alltt}
   ModelLoad(\textrm{\textit{Params}}, \textrm{\textit{Fields}}, \textrm{\textit{Vevs}}, \textrm{\textit{W}}, \textrm{\textit{V\_D}}, \textrm{\textit{Fterms}}, \textrm{\textit{mF}}, \textrm{\textit{mB}})
\end{alltt}
is called implicitly as the last line of any file which was created with \verb|ModelSave()|, so you will rarely need to use it explicitly. Like \verb|ModelConstruct()|, this command loads a specific model into \verb|Vscape| such that it can be studied with subsequent commands. However, in this case the F-terms and mass matrices need to be specified as an argument, so this command avoids the sometimes time-consuming computation of the F-terms and mass matrices. \textit{Fterms} should be a list of expressions for the different F-terms. \textit{mF} and \textit{mB} are the fermion and scalar mass matrices respectively. The entries in \textit{Fterms}, \textit{mF} and \textit{mB} should be ordered as the fields are ordered in \textit{Fields}. Thus e.g. the second entry in the list \textit{Fterms} is the F-term with respect to the second field appearing in the list \textit{Fields}. Warning: the program does not check whether the given F-terms and mass matrices actually correspond to the superpotential given!

If you invoke \verb|ModelConstruct()| (or implicitly \verb|ModelLoad()| via \verb|importfile()|), when a model was already loaded, the internal model of Vscape will be changed to the new model. Thus, all the subsequent commands will involve that new model. 

\paragraph{Example}
This example and the subsequent examples use the model discussed in \cite{ISS1} with $N = 1$ and $N_f = 6$. A slightly modified version of this example is included in the distribution of Vscape and can be loaded with \verb|importfile("|\textit{path/}\verb|examples/iss.txt")| where \textit{path} is the correct path to the file. Note: the file \verb|iss.txt| was not created with \verb|ModelSave()|, the file does not contain F-terms nor mass matrices, so you savely modify it.
\begin{verbatim}
   > ## Switch to real mode, to introduce two real parameters
   > real_symbols;
     All new symbols introduced from now on are assumed to be real.
   > Lambda:
   > m_ISS:
   >
   > ## Switch to complex mode, to introduce all the other symbols
   > complex_symbols;
     All new symbols introduced from now on are assumed to be complex.
   >  
   > ## Define the model
   > Params = {Lambda, m_ISS}:
   > B  = symbolicmatrix(6,1,B_);
   [[B_0],[B_1],[B_2],[B_3],[B_4],[B_5]]
   > Bt = symbolicmatrix(1,6,Bt_):
   > M  = symbolicmatrix(6,6,M_):
   > Fields = {B,Bt,M}:
   > Vevs   = {B_5,Bt_5,M_00,M_11,M_22,M_33,M_44,M_55}:
   > W   = evalm(Bt*M*B)                //
   >       - determinant(M)*Lambda^(-3) //
   >       + Lambda*m_ISS*trace(M):
   > V_D = 0:
   >
   > ## Construct the model
   > ModelConstruct(Params, Fields, Vevs, W, V_D);
   Loading Model...
     Computing F-terms and mass matrices...
     F-terms:       i = 047;
     Mass matrices: i = 047; j = 047; k = 047;
     Done.
   Model successfully loaded...
   0
   > 
   > ## Now we save the model to a file
   > ModelSave("examples/isscomputed.txt");
   The internal model was successfully saved to <examples/isscomputed.txt>.
   0
\end{verbatim}
From now on this specific model can quickly be loaded using the command \verb|importfile("examples/isscomputed.txt");|. Note that the file \verb|isscomputed.txt| was created using \verb|ModelSave()| and thus contains the correctly ordered F-terms and mass matrices of this specific model, so this file should not be modified by hand. Whenever you want to define a new model different from the above one (e.g. with one more term in the superpotential) you will have to use the command \verb|ModelConstruct()| again)

\subsection{Symbolic commands for the model} \label{SymbolicCommands}
There are several commands which give you symbolic information on the model that is loaded. 
\begin{alltt}
   ModelParams()
      \textrm{Returns the parameters of the model in a list}   
   ModelFields()
      \textrm{Returns the fields of the model in a list}   
   ModelVevs()
      \textrm{Returns the fields with non-zero expectation value in a list}   
   ModelW()
      \textrm{Returns the superpotential}
   ModelVD()
      \textrm{Returns the D-term potential}
   ModelV0()
      \textrm{Returns the tree level potential symbolically}
   ModelFtermsi(\textrm{\textit{integer}})
   	  \textrm{Returns the F-terms related to field \textit{integer}}
   ModelFterms()      
   	  \textrm{Returns the F-terms in a list}
   ModelmFij(\textrm{\textit{a}}, \textrm{\textit{b}})    
      \textrm{Returns entry \textit{a}, \textit{b} of the fermion mass matrix: \(W_{ab}\)}
   ModelmBij(\textrm{\textit{a}}, \textrm{\textit{b}})   
      \textrm{Returns entry \textit{a}, \textit{b} of the upper right block of the scalar mass matrix: \(W*_{abc} W_c\)}
   ModelmF()   
      \textrm{Returns the fermion mass matrix: \(W_{ab}\)}
   ModelmB()          
      \textrm{Returns the upper right (off diagonal) block of the scalar mass matrix: \(W*_{abc} W_c\)}
\end{alltt}
The list of F-terms and the entries in the mass matrices are ordered according to the occurence of the fields in \verb|ModelFields()|. The \textit{integer} values in the above command also refer to the occurence of the field in \verb|ModelFields()|, the first field has integer \verb|0|.

\paragraph{Example}
To try the example below, you will need the file \verb|examples/isscomputed.txt| which was created in the example above.
\begin{verbatim}
   > importfile("examples/isscomputed.txt");
   > ModelFields();
   {B_0,B_1,B_2,B_3,B_4,B_5,
    Bt_0,Bt_1,Bt_2,Bt_3,Bt_4,Bt_5,
    M_00,M_01,M_02,M_03,M_04,M_05,
    M_10,M_11,M_12,M_13,M_14,M_15,
    M_20,M_21,M_22,M_23,M_24,M_25,
    M_30,M_31,M_32,M_33,M_34,M_35,
    M_40,M_41,M_42,M_43,M_44,M_45,
    M_50,M_51,M_52,M_53,M_54,M_55}
   > ModelFterms();
   {0,0,0,0,0,Bt_5*M_55,
    0,0,0,0,0,M_55*B_5,
    m_ISS*Lambda-M_33*Lambda^(-3)*M_44*M_11*M_55*M_22,0,0,0,0,0,
    0,m_ISS*Lambda-M_33*M_00*Lambda^(-3)*M_44*M_55*M_22,0,0,0,0,
    0,0,-M_33*M_00*Lambda^(-3)*M_44*M_11*M_55+m_ISS*Lambda,0,0,0,
    0,0,0,-M_00*Lambda^(-3)*M_44*M_11*M_55*M_22+m_ISS*Lambda,0,0,
    0,0,0,0,m_ISS*Lambda-M_33*M_00*Lambda^(-3)*M_11*M_55*M_22,0,
    0,0,0,0,0,-M_33*M_00*Lambda^(-3)*M_44*M_11*M_22+Bt_5*B_5+m_ISS*Lambda}   
   > 
   > ## F-term wrt B_5;
   > myFterm = ModelFtermsi(5);    
   Bt_5*M_55 
   >
   > ## entry associated with B_0, Bt_0 in mass matrices; 
   > ModelmFij(0,6); 
   M_00
   > ModelmBij(0,6);
   m_ISS*Lambda-M_33*Lambda^(-3)*M_44*M_11*M_55*M_22
\end{verbatim}

\subsection{Numerical commands for the model: the effective potential}
All commands that involve numeric computations work with double precision, independent from the \verb|Digits| setting for the symbolic computations. The commands to evaluate the potential numerically are,
\begin{alltt}
   ModelnumV0(\textrm{\textit{SubsList}})
      \textrm{Evaluates the tree level potential numerically}   
   ModelnumVcw(\textrm{\textit{SubsList}})
      \textrm{Evaluates the Coleman-Weinberg potential numerically}   
   ModelnumVeff(\textrm{\textit{SubsList}})
      \textrm{Evaluates the one loop effective potentiel potential numerically}   
   ModelnumSTrM2(\textrm{\textit{SubsList}})
      \textrm{Evaluates the supertrace over scalar and fermion mass matrices}
\end{alltt}
The \textit{SubsList} is a list of equations which specify the value of the parameters and fields which were allowed to have non-zero expectation value. 

\paragraph{Example}
The example below uses the file \verb|examples/isscomputed.txt| which was created in an example above.
\begin{verbatim}
   > importfile("examples/isscomputed.txt");
   > Subs = {Lambda == 1000.0, m_ISS == 10.0,  //
   >         B_5  == 0.00 + 100.0*I,           //
   >         Bt_5 == 0.01 + 100.0*I,           //
   >         M_00 == 0.0,                      //
   >         M_11 == 0.0,                      //
   >         M_22 == 0.0,                      //
   >         M_33 == 0.0,                      //
   >         M_44 == 0.0,                      //
   >         M_55 == 0.0}:
   > ModelnumVcw(Subs);
   3.3551986738766286522E7
   > ModelnumSTrM2(Subs);
   3.456079866737127304E-11
\end{verbatim}

\subsection{Minima of the effective potential}
You can find local minima of the one loop effective potential by using the command:
\begin{alltt}
   ModelMinSimpl(\textrm{\textit{SubsPattern}}, \textrm{\textit{Variables}}, \textrm{\textit{StartPoint}})
\end{alltt}
\textit{SubsPattern} is a list of equations which specifies the values of the parameters and fields with non-zero vev as function of the variables. The variables are specified in the field \textit{Variables} as a list of real symbols. The minimization algorithm will minimize the effective potential as a function of those variables. The field \textit{StartPoint} is a list of initial value for each of the variables. The command returns a list with values for the variables at the minima. The value for simplex size shown while the command is running is discussed in section \ref{MinSimplPrecision}

This command is calculationally intensive, the calculation time depends on the number of fields in the model. While the calculation progresses the computer will display the current iteration of the algorithm and the minimum value of the potential that has been reached so far. The algorithm used to minimize the function is the simplex algorithm of Nelder and Mead \cite{NelderMead}. This algorithm has the advantage that it does not rely on computational intensive gradients to iterate towards the minimum. Some more details and parameters that influence the simplex algorithm are given in section \ref{MinSimplPrecision}.

Another command that analyzes the effective potential is
\begin{alltt}
   ModelStatBox(\textrm{\textit{SubsPattern}}, \textrm{\textit{Variables}}, \textrm{\textit{StartPoint}}, \textrm{\textit{Point1}}, \textrm{\textit{Point2}})
\end{alltt}
where the fields \textit{SubsPattern} and \textit{Variables} have the same function described above. The command starts from the value of the effective potential where the variables have the value specified in \textit{StartPoint}. The algorithm then picks several random points in the box defined by \textit{Point1} and \textit{Point2} and checks whether the value of the effective potential at that point is lower than the minimum value found so far. The command returns a list of values for the variables at the random point with lowest effective potential.

\paragraph{Example}
The example below uses the file \verb|examples/isscomputed.txt| which was created in an example above.
\begin{verbatim}
   > importfile("examples/isscomputed.txt");
   > Subs = {Lambda == 1000.0, m_ISS == 10.0, //
   >         B_5  == 0 + b*I,                  //
   >         Bt_5 == c + d*I,                  //
   >         M_00 == 0.0,                      //
   >         M_11 == 0.0,                      //
   >         M_22 == 0.0,                      //
   >         M_33 == 0.0,                      //
   >         M_44 == 0.0,                      //
   >         M_55 == e}:
   > ModelMinSimpl(Subs, {b,c,d,e}, {100,0,100,0});
          250:  Veff = +5.33497644174886226654e+08   
                Simplex size = +1.56969168925476587682e-05
   {98.755419109029958236, 8.6554575838203248834E-7,
    98.755388956674764245, 7.9209619553734771633E-7}
   > ModelMinSimpl(Subs, {b,c,d,e}, {0,900,900,900});
          680:  Veff = +5.33497644174886107445e+08
                Simplex size = +3.02342927540789435954e-06
   {98.7554136717079416,  -3.5353383568610648586E-8,
    98.755391871241812396, 5.7324975688985215976E-8}
   >
   >
   > Subs = {Lambda == 1000.0, m_ISS == 10.0, //
   >         B_5  == 0 + b*I,                  //
   >         Bt_5 == c + d*I,                  //
   >         M_00 == a,                        //
   >         M_11 == a,                        //
   >         M_22 == a,                        //
   >         M_33 == a,                        //
   >         M_44 == a,                        //
   >         M_55 == e}:
   > ModelMinSimpl(Subs,{a,b,c,d,e},{900,0,0,0,900});
          660:  Veff = -1.03509432378173447649e-05
                Simplex size = +1.69541545589374184433e-13
   {398.10717886889341344,
      5.7428257024330178013E-7, 
      1.7617897585483444842E-6,
     -8.3947951601006729255E-7, 
    398.107133987191105}
   >
   > ModelStatBox(Subs,{a,b,c,d,e},{0,98,0,98,0}, //
   >              {200,0,0,0,200},{500,0,0,0,500});
      Checking random point no.    90; (so far     6 lower points found)
   {381.46924413740634918, 0.0, 0.0, 0.0, 467.38906654063612223} 
\end{verbatim}
The first two minimizations started at different points but led to the same metastable minimum discussed in \cite{ISS1}. Both results are equal up to numerical precision as discussed in section \ref{PrecisionVevs}. The $U(1)_B$ symmetry is spontaneously broken giving rise to a Goldstone direction. We fixed this freedom by choosing $\mathrm{Re} B_5 = 0$. 

The third minimization led to the supersymmetric vacua of the model. Notice that the rank-condition mechanism for supersymmetry breaking for models with $N = 1$ only holds around the origin in fieldspace. Away from the origin there is a point where all the F-terms become zero due to the additional determinant term in the superpotential. The numerical value for the effective potential at the supersymmetric point is slightly negative but zero within error bounds as discussed in section \ref{PrecisionCW}.

The fourth operation was a statistical search. We started from a point close to the metastable vacuum and search randomly in the interval $M_{00} = \cdots = M_{44} \in [200, 500], M_{55} \in [200, 500]$. The algorithm returned a point in fieldspace relatively close to the supersymmetric minima with a lower value for the effective potential.

\subsection{Analyzing the vacua}
The commands to analyse a minimum all use the same structure of arguments as described above: \textit{SubsPattern} is a list of equations linking the parameters and fields with non-zero vev to the variables which are specified in the list \textit{Variables}. The values for the variables are given in the list of numbers \textit{Point}.
\begin{alltt}
   ModelGradient(\textrm{\textit{SubsPattern}}, \textrm{\textit{Variables}}, \textrm{\textit{Point}})
      \parbox[t]{\mylength}{\textrm{Evaluates the gradient of the effective potential at \textit{Point}. The command returns a list with the gradient followed by error flags on the gradient each formatted as a 1 by n matrix. The computer will show the estimated time to complete the algorithm since this computation is calculationally intensive.}}  
\vspace{.02in}   ModelHessian(\textrm{\textit{SubsPattern}}, \textrm{\textit{Variables}}, \textrm{\textit{Point}})
      \parbox[t]{\mylength}{\textrm{Evaluates the Hessian matrix at \textit{Point}. The command returns a list with the hessian matrix followed by the matrix of error flags. The computer will show the estimated time to complete the algorithm since this computation is calculationally intensive.}}
\vspace{.02in}   Modeldiff2(\textrm{\textit{SubsPattern}}, \textrm{\textit{Variables}}, \textrm{\textit{Point}}, \textrm{\textit{i}}, \textrm{\textit{j}}, \textrm{\textit{dx\_i}}, \textrm{\textit{dx\_j}})
      \parbox[t]{\mylength}{\textrm{Evaluates the second order derivative with respect to fields \textit{i} and \textit{j}. \textit{dx\_i} and \textit{dx\_j} are suggested stepsize to compute the derivative (see section \ref{DiffPrecision} for more information). The command returns a list with the second order derivative followed by the error flag.}}
\vspace{.02in}   ModelNegEig(\textrm{\textit{SubsPattern}}, \textrm{\textit{Variables}}, \textrm{\textit{Point}}, \textrm{\textit{Hessian}}, \textrm{\textit{HessianError}})
      \parbox[t]{\mylength}{\textrm{Checks whether there are negative eigenvalues of the Hessian matrix of the effective potential at \textit{Point}. You need to make sure that \textit{Hessian} (and \textit{HessianError}) are indeed the Hessian matrix (and error flags) of the effective potential at \textit{Point}. The function will return the original \textit{Point} or, if there are negative eigenvalues that are larger than the average errorflag, another \textit{Point} where the effective potential has a lower value. Used as cross check for the other minimization commands.}}  
\end{alltt}

\paragraph{Example}
The example below uses the file \verb|examples/isscomputed.txt| which was created in an example above.
\begin{verbatim}
   > importfile("examples/isscomputed.txt");
   > Subs = {Lambda == 1000.0, m_ISS == 10.0, //
   >         B_5  == 0.0 + b*I,                //
   >         Bt_5 == 0.0 + d*I,                //
   >         M_00 == 0.0,                      //
   >         M_11 == 0.0,                      //
   >         M_22 == 0.0,                      //
   >         M_33 == 0.0,                      //
   >         M_44 == 0.0,                      //
   >         M_55 == 0.0}:
   > Min = ModelMinSimpl(Subs, {b,d}, {100,100});
          160:  Veff = +5.33497644174886047840e+08
                Simplex size = +3.29244092296298119244e-06
   {98.75539482502895794,98.75540931029662772}
   >
   >  
   > ModelGradient(Subs, {b,d}, {100,100});
         2/2:  00:01s to go...
   {[[44055.476986700297857, 44055.476986670022598]],
    [[6.273248519121328748,      6.27324852896071139]]}  
   > ModelGradient(Subs, {b,d}, Min);
         2/2:  00:01s to go...
   {[[-5.261715075073619019,    -5.2542854259666373906]],
    [[ 3.8105577427291033032,    3.81055711846567835]]}     
   >   
   > ModelHessian(Subs, {b,d}, Min);
         3/3:  00:19s to go...
   {[[17652.30720837064655,     17140.548392404118204],
     [17140.548392404118204,    17652.301781115671474]],
    [[   10.064568391628851529,    10.455072325682765566],
     [   10.455072325682765566,    10.064565316884026558]]}   
   > Hessian = op(%, 0): HessianErr = op(%%, 1):
   > eigensymm(Hessian);
   {[[34792.852887147491856,      511.75610233882599687]],
    [[    0.7071068371597952762,   -0.70710672521329531737],
     [    0.70710672521329531737,   0.7071068371597952762]]}
\end{verbatim}
The second column in the last matrix in the example is the eigenvector corresponding to the pseudo-moduli direction in \cite{ISS1}. The corresponding mass squared eigenvalue indeed corresponds to the theoretical value computed in \cite{ISS1}: $511.756 \, \mathrm{GeV}^2$ versus $489.248 \, \mathrm{GeV}^2$.

\subsection{Plots}
There are three commands in Vscape to plot:
\begin{alltt}
   ModelPlotV0(\textrm{\textit{filename}}, \textrm{\textit{SubsPattern}}, \textrm{\textit{Variables}}, \textrm{\textit{xmin}}, \textrm{\textit{xmax}}, \textrm{\textit{ymin}}, \textrm{\textit{ymax}})
   ModelPlotVcw(\textrm{\textit{filename}}, \textrm{\textit{SubsPattern}}, \textrm{\textit{Variables}}, \textrm{\textit{xmin}}, \textrm{\textit{xmax}}, \textrm{\textit{ymin}}, \textrm{\textit{ymax}})
   ModelPlotVeff(\textrm{\textit{filename}}, \textrm{\textit{SubsPattern}}, \textrm{\textit{Variables}}, \textrm{\textit{xmin}}, \textrm{\textit{xmax}}, \textrm{\textit{ymin}}, \textrm{\textit{ymax}})
\end{alltt}
Each commands writes 3d plotting data of respectively $V_0$, $V_{\mathrm{cw}}$ or $V_{\mathrm{eff}}$ to the file \textit{filename}. \textit{SubsPattern} again indicates how the parameters and fields with non-zero vev depend on the \textit{Variables}. \textit{Variables} should be a list of 2 variables which will be on the x and y axis of the 3d plot. The variables will be evaluated between \textit{xmin} and \textit{xmax} (\textit{ymin}, \textit{ymax} resp.) with a stepsize of $1$. The computer will show the progress for all plotting commands since it is calculationally intensive. The command returns $0$ when it was successful. The file is formatted such that it can then easily be imported in other math packages. A maple file (\verb|Plotting.mw|) which can import the data is included in the distribution of Vscape in the directory \verb|examples/|. The file format is 
\begin{verbatim}
   xmin xmax
   ymin ymax                                                          
   f(xmin,ymin)   f(xmin+1,ymin)   ... f(xmax,ymin)
   f(xmin,ymin+1) f(xmin+1,ymin+1) ... f(xmax,ymin+1)
   ...                                                             
   f(xmin,ymax)   f(xmin+1,ymax)   ... f(xmax,ymax)
\end{verbatim} 
where the first two lines contain integer values, while all other data consists of doubles. 

\paragraph{Example}
The example below uses the file \verb|examples/isscomputed.txt| which was created in an example above.
\begin{verbatim}
   > importfile("examples/isscomputed.txt");
   > pi = evalf(Pi):
   > M  = 398.1*b/80:
   > B  = 98.755*(1 - b/80): 
   > Subs = {Lambda == 1000.0, m_ISS == 10.0,  //
   >        B_5  == exp( 2*pi*a*I/10)*(B*I),   //
   >        Bt_5 == exp(-2*pi*a*I/10)*(B*I),   //
   >        M_00 == M,                         //
   >        M_11 == M,                         //
   >        M_22 == M,                         //
   >        M_33 == M,                         //
   >        M_44 == M,                         //
   >        M_55 == M}:
   > ModelPlotVeff("BarrierPlot.txt", Subs, {a,b}, 0, 9, 0,99);
   The 3d plotting data was successfully saved to <BarrierPlot.txt>.
   0        
\end{verbatim}
The plot obtained from the example using \verb|examples/plotting.mw| in Maple 10 is shown in figure \ref{fig:BarrierPlotVeff}.

\FIGURE{
  \centering
  \includegraphics[scale=0.5]{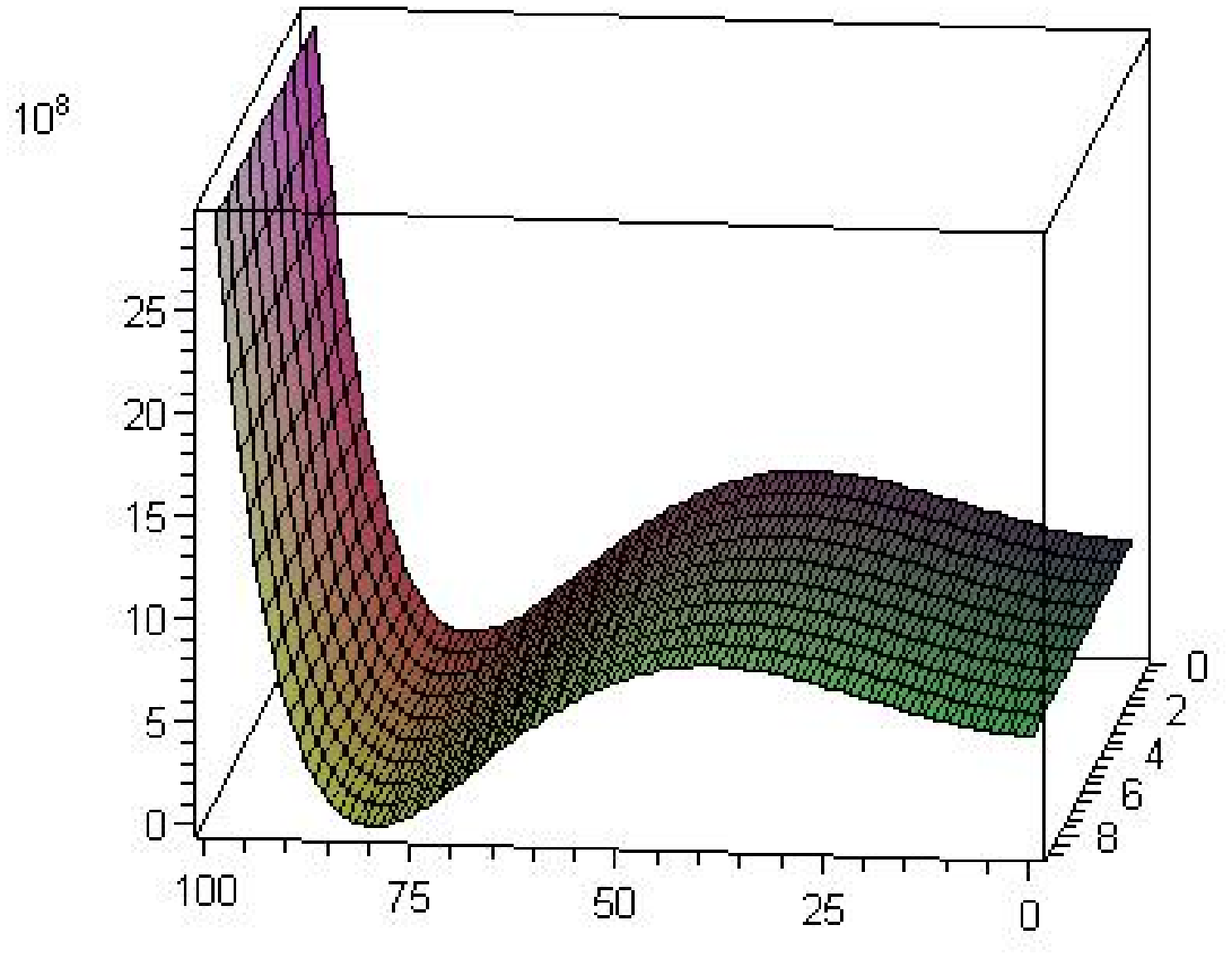}    
	\caption{The effective potential from the metastable vacuum to the supersymmetry vacuum. The axis into the paper is the flat Goldstone direction of the $U(1)_B$. The axis along the paper cuts through the one-loop effective potential from the metastable minima on the left to the supersymmetric vacuum on the right side.}
	\label{fig:BarrierPlotVeff}
} 

\subsection{Interface with spectrum generators}
The command 
\begin{alltt}
   ModelSpectr(\textrm{\textit{Filename}}, \textrm{\textit{SubsList}})
\end{alltt}
allows interaction with spectrum generators. The command takes the file \textit{filename} as input file. This file should be formatted according to the Les Houches Accord \cite{LesHouchesAccord}, except that numbers can be replaced by formulas. \verb|ModelSpectr()| will substitute the values specified in \textit{SubsList} and the current values of all other variables used up to that point in Vscape and generate an outputfile \verb|<filename.out>|. This file can then be used as input file for spectrum generators. Whenever an unknown symbol is encountered, the entire line is copied verbatim without substitution and a message is printed on the screen. Vscape will add a Block containing version information to the output file, to indicate that the file was generated with Vscape.

\paragraph{Example}
The example below uses the file \verb|examples/isscomputed.txt| which was created in an example above. As input file in this example, we use \verb|LesHoucheSymbolic.txt|
\begin{verbatim}
   # SUSY Les Houches Accord 1.0 - example input file
   #
   Block MODSEL                # Select Model
        1    2                 #   (m)GMSB
   Block SMINPUTS              # Standard Model inputs
        3    0.1172            #   alpha_s(MZ) SM MSbar
   Block MINPAR                # Susy breaking input parameters
        1  sqrt(Lambda*m_ISS)  #   scale of soft susy breaking
        3  ReHu0/ReHd0         #   tanb
        4    1.0               #   sign(mu)
   Block EXPAR                 # Non-minimal susy breaking input parameters
        23   g_S*S             #   mu-term
\end{verbatim}
The commands sequence of the example is
\begin{verbatim}
   > importfile("examples/isscomputed.txt");
   > Subs = {Lambda == 1000.0, m_ISS == 10.0, //
   >         B_5  == 0 + b*I,                  //
   >         Bt_5 == c + d*I,                  //
   >         M_00 == 0.0,                      //
   >         M_11 == 0.0,                      //
   >         M_22 == 0.0,                      //
   >         M_33 == 0.0,                      //
   >         M_44 == 0.0,                      //
   >         M_55 == e}:
   > ModelMinSimpl(Subs, {b,c,d,e}, {100,0,100,0});
          250:  Veff = +5.33497644174886226654e+08
                Simplex size = +1.56969168925476587682e-05
   {98.755419109029958236, 8.6554575838203248834E-7,
    98.755388956674764245, 7.9209619553734771633E-7}
   > MinSubs = subs(Subs, {b,c,d,e}, %);
   {Lambda==1000.0, m_ISS==10.0, 
    B_5==98.755419109029958236*I, 
    Bt_5==8.6554575838203248834E-7+98.755388956674764245*I, 
    M_00==0.0, M_11==0.0, M_22==0.0, M_33==0.0, M_44==0.0, 
    M_55==7.9209619553734771633E-7}
   > g_S = 0.6:
   > S   = 0.001:
   > ModelSpectr("LesHouchesSymbolic.txt", MinSubs);
   unknown symbol 'ReHu0'
    => just copied this data line from inputfile to outputfile...
   Unable to match all symbols in <LesHouchesSymbolic.txt>
   with currently defined symbols...
   2   
\end{verbatim}
This example produces the output file \verb|<LesHouchesSymbolic.txt.out>| 
\begin{verbatim}
   # SUSY Les Houches Accord 1.0 - example input file
   #
   Block VSCAPE          # program info
        1     Vscape     # program
        2     1.0.0      # version number
   Block MODSEL                # Select Model
        1   2                      #   (m)GMSB
   Block SMINPUTS              # Standard Model inputs
        3   0.1172                 #   alpha_s(MZ) SM MSbar
   Block MINPAR                # Susy breaking input parameters
        1   100.0       #   scale of soft susy breaking
        3  ReHu0/ReHd0         #   tanb
        4   1.0                    #   sign(mu)
   Block EXPAR                 # Non-minimal susy breaking input parameters
        23   6.0E-4                  #   mu-term
\end{verbatim}
Notice that the line specifying $\tan \beta$ was not replaced by a number, since our simple example did not specify a value for $\mathrm{Re} H_u^0$.

\subsection{Numerical precision and errorflags}  \label{SubSectionPrecision}
The numerical format used for all the numerical computations in Vscape is double precision, which has 52 significant bits or approximately 16 decimal digits. 

\subsubsection{Precision of the Coleman-Weinberg potential} \label{PrecisionCW}
In one of the examples above, we found that the effective potential at the supersymmetric minima had a negative numerical value of order $-10^{-5}$. The value for the vevs of some of the fields at that point was around $400$. This implies that some of the squares of the mass squared eigenvalues of the mass matrices will be of the order $m^4 \approx 400^4 \approx 10^{11}$. Those squares of mass squared eigenvalues appear in the Coleman-Weinberg potential. As the internal numerical algorithm uses double precision, we can expect the absolute error on the potential to be estimated by
\begin{equation}
  |\phi_{\mathrm{max}}|^4 10^{-16} \, .
\end{equation}
Or, in our example of the order of $10^{-5}$, which implies that the value for the effective potential at the supersymmetric minimum is zero within error bounds. 

\subsubsection{Precision of the vevs after minimization} \label{PrecisionVevs}
In another example, we found two slightly different sets of vevs for the fields at the metastable minima, since we started the minimization process from a different point. The relative difference between the vevs was of order $10^{-7}$. The vevs of some of the fields at the metastable minima was of order $10^2$. Following the reasoning from the previous paragraph, we estimate that the absolute error on the effective potential is of order $10^{-8}$ and since the actual value of the effective potential at the minimum was of order $10^8$, the relative error is of order $10^{-16}$. Minimizing the potential then holds,
\begin{eqnarray}
  V_{\mathrm{eff}}(\phi_i)   
      & \approx & V_{\mathrm{eff}}(\phi_i) \pm |\phi_{\mathrm{max}}|^4 10^{-16}   \\
      &     =   & V_{\mathrm{eff}}(\phi_i) (1 \pm 10^{-16})                       \\
      &     =   & V_{\mathrm{eff}}(\phi_i + \epsilon_i)                           \\
      &     =   &   V_{\mathrm{eff}}(\phi_i)
                  + \frac{1}{2} 
                    \frac{\partial^2 V_{\mathrm{eff}}}{\partial \phi_i \partial \phi_j} 
                    \epsilon_i \epsilon_j
\end{eqnarray}  
where $\epsilon_i$ is the error on the vev of the field $\phi_i$ at the minimum coming from the limited precision on $V_{\mathrm{eff}}$. In the example above, $V_{\mathrm{eff}}$ was of order $10^8$ at the metastable minima, while the second order derivative can be at most of order the cutoff squared, which was $10^6$. We find that the highest precision on the fields at the metastable minimum is of order $10^{-7}$, which corresponds to the results found in the example. A lower value for the second order derivative gives less precision.

\subsection{Parameters controlling the internal algorithms} \label{SubSectionControlParameters}
\subsubsection{Simplex minimization} \label{MinSimplPrecision}
The algorithm that is used for the command \verb|ModelMinSimpl()| is the Simplex algorithm by Nelder and Mead \cite{NelderMead}. An n-simplex is chosen around the startpoint in variable space. By changing the vertices of the simplex, the algorithm minimizes the size of the simplex step by step. When the size of the simplex is smaller than \textit{MinimizerSizeTest}, the algorithm considers the minimum to be found. When the command \verb|ModelMinSimpl()| is running, the current size of the simplex is shown on the screen. The parameter \textit{MinimizerInitialStep} determines the size of the initial simplex with which the algorithm starts. \textit{MinimizerMaxIterations} sets the maximal number of iterations, the algorithm will quit even if the algorithm has not found a minimum. Sometimes the size of the simplex never becomes small enough to get a succesfull end of the algorithm, while the simplex is already at its most optimal position. The parameter \textit{MinimizerMaxDeadIterations} determines the maximal number of iterations that the algorithm can proceed without finding a point with a lower functional value.

\subsubsection{Statistical minimization}
The number of random points the command \verb|ModelStatBox()| test for a lower value of the effective potential, is determined by the parameter \textit{StatBoxMaxIterations}.

\subsubsection{Derivatives} \label{DiffPrecision}
The numerical derivatives are computed using the 11-point rule:
\begin{eqnarray}
  \frac{df(x_0)}{dx} & \approx & \frac{1}{2520 \Delta x} 
                                                   \Big(-      2 f(x_0 - 5 \Delta x) 
                                                        +     25 f(x_0 - 4 \Delta x) 
                                                        -    150 f(x_0 - 3 \Delta x)  \nonumber \\
                     &   & \qquad \qquad \;             + \, 600 f(x_0 - 2 \Delta x)
                                                        -   2100 f(x_0 - 1 \Delta x)  \nonumber \\
                     &   & \qquad \qquad \;             + \,2100 f(x_0 + 1 \Delta x) 
                                                        -    600 f(x_0 + 2 \Delta x)  \nonumber \\
                     &   & \qquad \qquad \;             + \, 150 f(x_0 + 3 \Delta x) 
                                                        -     25 f(x_0 + 4 \Delta x)
                                                        +      2 f(x_0 + 5 \Delta x) \Big)  
\end{eqnarray}
where $\Delta x$ is the stepsize of the algorithm. The algorithm estimates the errorflags coming from rounding errors where the relative error on the effective potential is set by \textit{diffErrorOnf}. In addition, the truncation error which is estimated by comparing the 9-point rule with the 11-point rule result is also determined. The rounding error scales as $\Delta x^{-1}$ while the truncation error goes like $\Delta x^8$. Using those two competing errors, the algorithm estimates a new optimal $\Delta x$ and repeats the process until the relative error is smaller than \textit{diffTolerance} \cite{Recipes}. It will iterate at most \textit{diffMaxCycle} times.
One more parameter determines the behaviour of the algorithm, if the new guess for an optimal $\Delta x$ is bigger than \textit{diffMaxdx} then the algorithm quits. The command \verb|Modeldiff2()| allows you to suggest the initial stepsizes of the derivatives in the $i$ and $j$ direction with the arguments \textit{dx\_i}, \textit{dx\_j}. The other commands make an automatic initial estimate.

\subsubsection{Changing the control parameters}
You can get the values of all the parameters discussed above with the command
\begin{alltt}
   GetPrecision()
\end{alltt}
You can set the parameters with the command
\begin{alltt}
   SetPrecision(\textrm{\textit{StatBoxMaxIterations}}, \textrm{\textit{MinimizerSizeTest}}, \textrm{\textit{MinimizerInitialStep}}, \textrm{\textit{MinimizerMaxIterations}}, \textrm{\textit{MinimizerMaxDeadIterations}}, \textrm{\textit{diffMaxCycle}}, \textrm{\textit{diffTolerance}}, \textrm{\textit{diffMaxdx}}, \textrm{\textit{diffErrorOnf}})
\end{alltt}
If you set an argument in \verb|SetPrecision()| to zero, the parameter will maintain its old value.

The default parameters of the program have been optimized for models of the kind used in the examples, with a cutoff of the order of $1000$ where we think in units of $\mathrm{GeV}$. Different models might require different parameters. For example, if one has a very shallow metastable minima, \textit{MinimizerInitialStep} should not be chosen too large, since this would result in the algorithm jumping out of the metastable minima in favour of another minima with a lower effective potential.


\end{document}